\documentclass{elsart}
\usepackage{amsmath,amssymb,graphicx}

\renewcommand{\vec}[1]{\mbox{\boldmath$#1$}}

\def\mi{\mathrm{i}}

\begin{document}

\begin{frontmatter}

\title{Interaction for  Solitary Waves  with a Phase Difference in a Nonlinear Dirac Model}

\thanks[label2]{Corresponding author. Tel:\ +86-10-62757018; Fax:\ +86-10-62751801.}
\author{Sihong Shao\thanksref{label2}},
\ead{shaosihong@pku.edu.cn}
\author{Huazhong Tang}
\ead{hztang@pku.edu.cn}
\address{LMAM, School of Mathematical Sciences,\\
  Peking University, Beijing 100871, P.R. China.}

\begin{abstract}
This paper presents a further numerical study of the interaction
dynamics for  solitary waves in a nonlinear Dirac field with
scalar self-interaction by using a fourth order accurate
Runge-Kutta discontinuous Galerkin  method. Our experiments are
conducted on the Dirac solitary waves
 with a phase difference. Some interesting
 phenomena are observed: (a) full repulsion in binary and ternary collisions
and its dependence on the distance between initial waves; (b)
repulsing first, attracting afterwards, and then collapse in
binary and ternary collisions of initially resting two-humped
waves; (c) one-overlap interaction and two-overlap interaction in
ternary collisions of initially resting waves.
\end{abstract}

\begin{keyword}
Runge-Kutta discontinuous Galerkin method,  Dirac model,  solitary waves, 
phase difference, interaction dynamics.

\end{keyword}
\end{frontmatter}

\section{Introduction}
Ever since its invention in 1929 the Dirac equation has played a
fundamental role in various areas of modern physics and
mathematics, and is important for the description of interacting
particles and fields.
 Consider a classical spinorial model with scalar self-interaction, described by the nonlinear Lagrangian $ L = \mi
\overline{\vec{\psi}}\gamma^\mu\partial_\mu\vec{\psi}
-m\overline{\vec{\psi}}\vec{\psi}+\lambda(\overline{\vec{\psi}}\vec{\psi})^2
$ from which we may derive the nonlinear Dirac equation
\begin{equation}\label{eq::dirac1}
\mi \gamma^\mu\partial_\mu\vec{\psi}-m\vec{\psi} +
2\lambda(\overline{\vec{\psi}}\vec{\psi})\vec{\psi} = 0,
\end{equation}
 where the $\gamma^\mu$ matrices are defined by
\begin{equation*}
\gamma^0=\left(
\begin{matrix}
I & 0 \\
0 & -I
\end{matrix}
 \right),
\gamma^k=\left(
\begin{matrix}
0 & \sigma^k \\
-\sigma^k & 0
\end{matrix}
 \right),
\end{equation*}
here $\sigma^k$ with $k=1,2,3$, denote the Pauli matrices. The
nonlinear self-coupling term $(\overline{\vec{\psi}}\vec{\psi})^2$
in  the Lagrangian allows the existence of finite energy, localized
solitary waves, or extended particle-like solutions, see e.g.
\cite{soler1970}. Several authors have committed themselves to
analytically investigating  the nonlinear Dirac model
\cite{rs1973,rrsv1974,alvarez1983ene,as1986,alvarez1985,alvarez1988}.

Here we only pay our attention to advances in numerical studies of
interaction dynamics of the Dirac solitary waves. Up to now, some
reliable, higher-order accurate numerical methods have been
constructed to solve the nonlinear Dirac equation
(\ref{eq::dirac1}). They include  Crank-Nicholson type schemes
\cite{alvarez1981,alvarez1992}, split-step spectral
schemes \cite{sss}, Legendre rational spectral methods
\cite{wg2004},
and Runge-Kutta discontinuous Galerkin (RKDG) methods
\cite{shao-tang01}, etc. The interaction dynamics for  the solitary
wave solutions  of (\ref{eq::dirac1}) were numerically simulated in
\cite{alvarez1981}
 by using a second-order accurate difference scheme. The authors 
 saw  there:
 charge and energy interchange except for some particular initial velocities of the solitary
waves; inelastic interaction in  binary collisions; and bound state production from binary collisions.
Weakly  inelastic interaction in  ternary collisions
is observed in \cite{shao-tang01}.
 The interaction dynamics in the
binary and ternary collisions of  two-humped solitary waves  are first
investigated in \cite{shao-tang02}.

However,  the experiments  carried out in the literatures are all
limited to the binary and ternary collisions of  the in-phase
solitary waves of  (\ref{eq::dirac1}). In this Letter we will
devote ourselves to further investigating the interaction dynamics
in the binary and ternary collisions of the Dirac solitary waves
with an initial phase shift
 by using a fourth-order accurate RKDG method \cite{shao-tang01}
 and report some interesting observations. The
RKDG methods adopt a discontinuous piecewise polynomial space for
the approximate solutions and the test functions, and an explicit,
high-order Runge-Kutta time discretization. It has been demonstrated by
various experiments that the fourth-order RKDG  method is numerically stable
without generating numerical oscillation within a very long time
interval,  has uniformly numerical convergence-rates, and preserve
conservation of the energy and charge.
 We refer the reader to \cite{shao-tang01} as well as \cite{shao-tang02} for a
detailed description and more numerical demonstrations.

\section{Preliminaries}

We restrict our attention to the $(1+1)$-dimensional
 nonlinear Dirac model (\ref{eq::dirac1}), and use the same notations $\rho_E(x,t)$ and $\rho_Q(x,t)$  as ones in
  \cite{shao-tang01,shao-tang02} to denote the energy and charge
  densities defined by
\begin{align}
\rho_E(x,t) =&\mbox{Im}(\psi_1^*\partial_x\psi_2+\psi_2^*\partial_x\psi_1)
+m(|\psi_1|^2-|\psi_2|^2)
-\lambda(|\psi_1|^2-|\psi_2|^2)^2,\label{eq::energy}\\
\rho_Q(x,t)  =& |\psi_1|^2+|\psi_2|^2,\label{eq::charge}
\end{align}
where $\psi_1$ and $\psi_2$ are two components of the spinor $\vec{\psi}(x,t)$.
A standing wave solution of the Dirac model (\ref{eq::dirac1}) is given as
\begin{eqnarray}\label{sol:sw}
\vec{\psi}^{sw}(x,t) \equiv \left(
\begin{array}{l}
\psi_1^{sw}(x,t)\\ \psi_2^{sw}(x,t)
\end{array}
\right) = \left(
\begin{array}{l}
A(x) \\ \mi B(x)
\end{array}
\right)e^{-\mi\Lambda t},\ 0<\Lambda\leq m,
\end{eqnarray}
with
\begin{eqnarray}
A(x)=\frac{\sqrt{\frac{1}{\lambda}(m^2-\Lambda^2)(m+\Lambda)}\cosh\big(x\sqrt{(m^2-\Lambda^2)}\big)}
{m+\Lambda\cosh\big(2x\sqrt{(m^2-\Lambda^2)}\big)},\\
B(x)=\frac{\sqrt{\frac{1}{\lambda}(m^2-\Lambda^2)(m-\Lambda)}\sinh\big(x\sqrt{(m^2-\Lambda^2)}\big)}
{m+\Lambda\cosh\big(2x\sqrt{(m^2-\Lambda^2)}\big)}.
\end{eqnarray}

The Dirac model (\ref{eq::dirac1}) also has  a single solitary
wave solution placed initially at $x_0$ with a velocity $v$:
\begin{equation}
\vec{\psi}^{ss}(x-x_0,t) =\big(\psi_1^{ss}(x-x_0,t),
\psi_2^{ss}(x-x_0,t) \big)^T,\label{sol:ss}
\end{equation}
where
\begin{align}
&\psi_1^{ss}(x-x_0,t) =
\sqrt{\frac{\gamma+1}{2}}\psi_{1}^{sw}(\tilde{x},\tilde{t})
+\mbox{sign}(v)\sqrt{\frac{\gamma-1}{2}}\psi_{2}^{sw}(\tilde{x},\tilde{t}),
\\
&\psi_2^{ss}(x-x_0,t) =\sqrt{\frac{\gamma+1}{2}}\psi_{2}^{sw}(\tilde{x},\tilde{t})+
\mbox{sign}(v)\sqrt{\frac{\gamma-1}{2}}\psi_{1}^{sw}(\tilde{x},\tilde{t}),
\end{align}
here $\gamma = {1}/{\sqrt{1-v^2}}$, $\tilde{x} =
\gamma(x-x_0-vt)$, $\tilde{t} = \gamma(t-v(x-x_0))$,
$\psi_{1}^{sw}$ and $\psi_{2}^{sw}$ are defined in (\ref{sol:sw})
and $\mbox{sign}(x)$ is the sign function, which returns $1$ if
$x>0$, $0$ if $x=0$, and $-1$ if $x<0$. The function
$\vec{\psi}^{ss}(x-x_0,t)$ represents a solitary wave travelling
from left to right if $v>0$, or  travelling from right to left if
$v<0$, and the standing wave $\vec{\psi}^{sw}(x-x_0,t)$ is
actually a solitary wave at rest placed at $x_0$ or identical to
$\vec{\psi}^{ss}(x-x_0,t)$ with $v=0$.
\begin{figure}
\centering
\includegraphics[width=6.6cm,height=5cm]{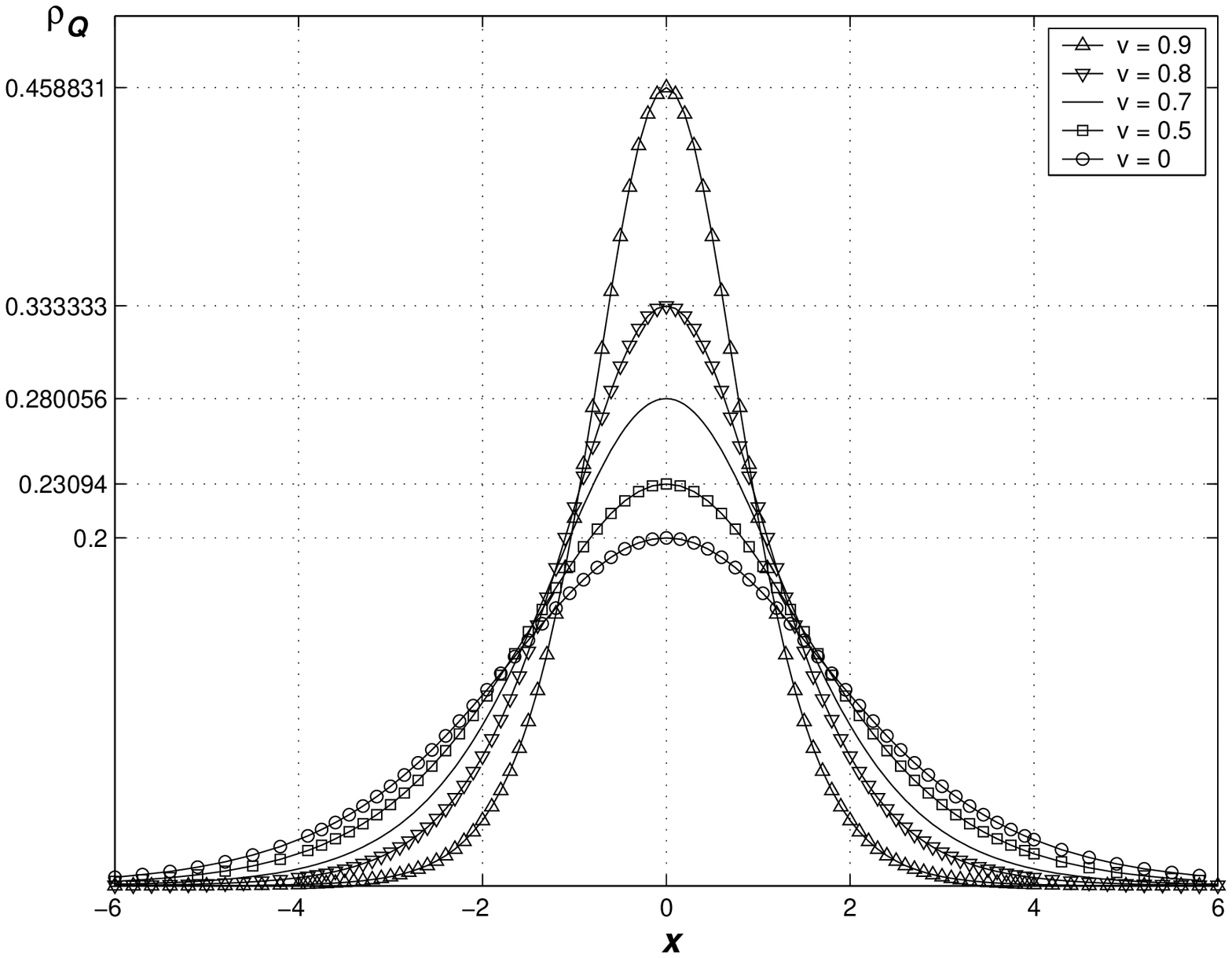}
\includegraphics[width=6.6cm,height=5cm]{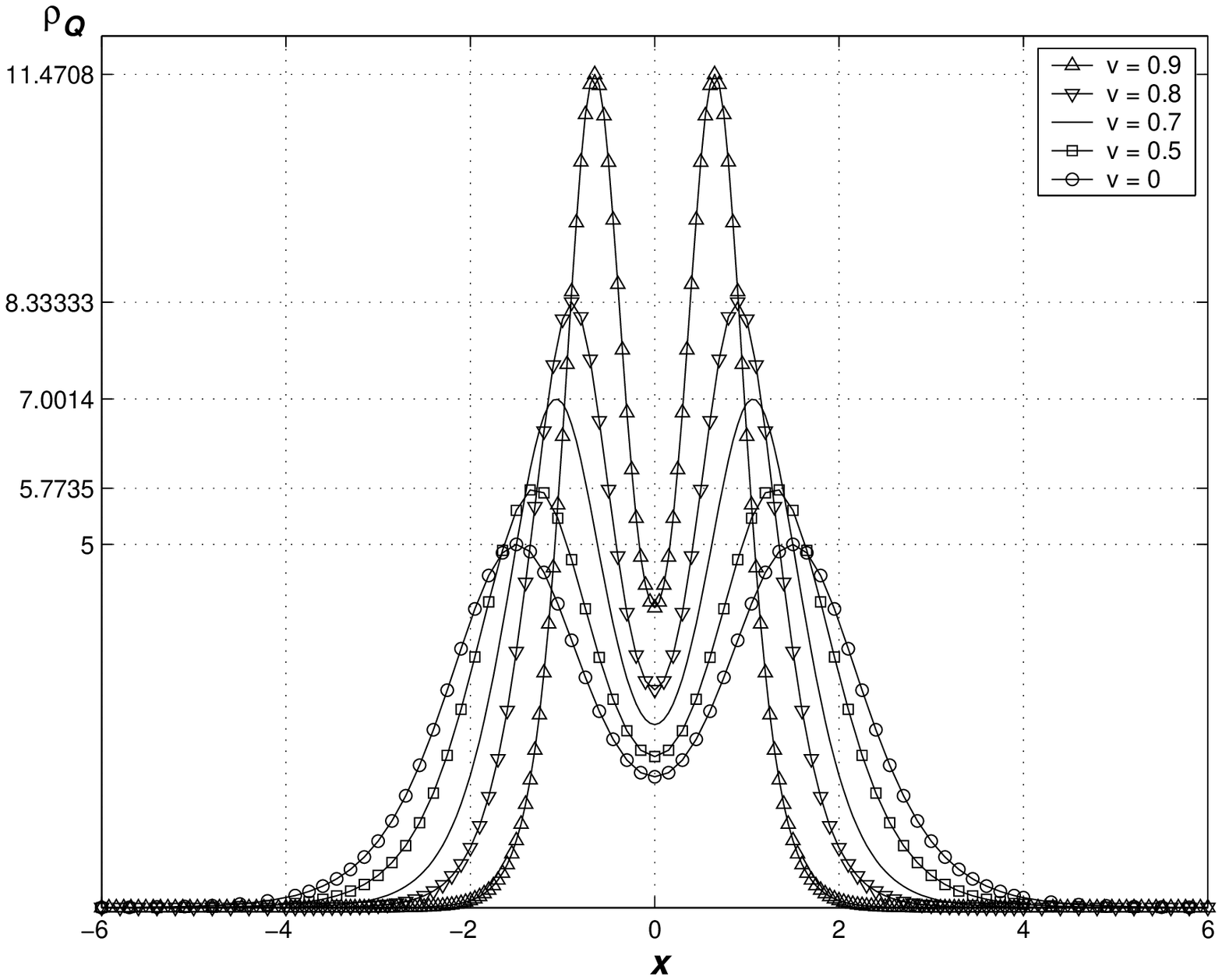}
\caption{\small Dependence of  $\rho_Q$ on $\Lambda$ and $v$. Left: $\Lambda = 0.9$;  right: $\Lambda = 0.1$.
 }\label{fig::Lambda_rhoq}
\end{figure}

 The profile of the solution (\ref{sol:sw}) or (\ref{sol:ss}) is strongly
dependent on the parameter $\Lambda$: it is a two-humped solitary
wave with two peaks whose locations are determined by
$\cosh(2\sqrt{m^2-\Lambda^2}\tilde{x})=\frac{m^2-\Lambda^2}{m\Lambda}$
 if $0<\Lambda<\frac{m}{2}$;
it becomes  a one-humped solitary wave with one peak located at
$\tilde{x}=0$ if $\frac{m}{2}\leq\Lambda<m$; and
$\vec{\psi}^{ss}(x-x_0,t)\equiv 0$ if $\Lambda=m$. Moreover,
amplitude of the solitary waves also depends strongly on the
velocity $v$: $\rho_Q^{ss}(x-x_0,t) =
\gamma\rho_Q^{sw}(\tilde{x},\tilde{t})$.
 Fig.\ \ref{fig::Lambda_rhoq} shows that dependence, which will gives different
 interaction dynamics.
 It is worth noting
  that  $e^{\mi\theta}\vec{\psi}^{ss}(x-x_0,t)$ is still
 a solitary wave solution of the Dirac model (\ref{eq::dirac1}), if $\theta$ is a
 constant.

In the following, our computations will work in dimensionless
units, or equivalently, take $m=1$ and $\lambda=\frac12$, and
adopt the non-reflecting boundary conditions at two boundaries of
the computational domain. The domain is covered by some identical
cells with area of $0.05$. The numerical algorithm is the
$P^3$-discontinuous Galerkin method in space combined with a
fourth order accurate Runge-Kutta time discretization, see
\cite{shao-tang01}.

\section{Binary collisions}

We solve (\ref{eq::dirac1}) with the initial data
\begin{equation}\label{eq:os1}
\vec{\psi}(x,0)=e^{\mi\theta_l}
\vec{\psi}^{ss}(x-x_l,0)+e^{\mi\theta_r}\vec{\psi}^{ss}(x-x_r,0),
\end{equation}
where $\theta_l$ and $\theta_r$ are two real numbers, determining
whether two waves are in phase. For convenience, we will say that
two solitary waves are equal if $\Lambda_l=\Lambda_r$ and
$|v_l|=|v_r|$. Throughout our numerical experiments in this
section, we will take $\theta_l=0$ and $\theta_r=\pi$, unless
stated otherwise.

\subsection{Two one-humped solitary waves}
\begin{figure}
\centering
\includegraphics[width=6.6cm,height=4.5cm]{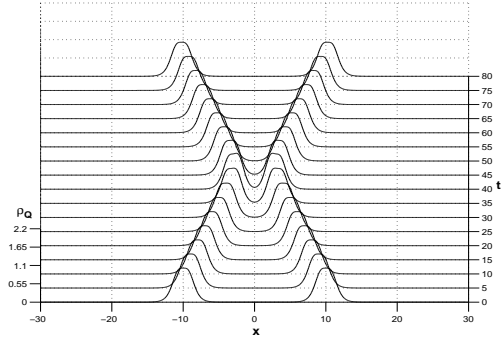}
\caption{\small  The time evolution of the charge density
$\rho_Q$. $\Lambda_l=\Lambda_r=0.5$, $v_l=-v_r=0.2$,
$x_r=-x_l=10$. 
}
\label{fig::caseB1}
\end{figure}

In this subsection, we study the interaction dynamics of two
one-humped solitary waves with a phase shift of $\pi$. The left
plot of Fig.\ \ref{fig::caseB1} shows the computed results for the
case of two equal one-humped solitary waves, i.e.
$\Lambda_l=\Lambda_r=0.5$, $v_l=-v_r=0.2$, and $x_r=-x_l=10$. We
see that the elastic interaction happens and two  one-humped
solitary waves  keep their initial shapes and velocities after
their collisions. It is worth noting that strong overlap happens
if above two initial waves are in phase, i.e.
$\theta_l=\theta_r=0$.%

\begin{figure}
\includegraphics[width=6.6cm,height=4.5cm]{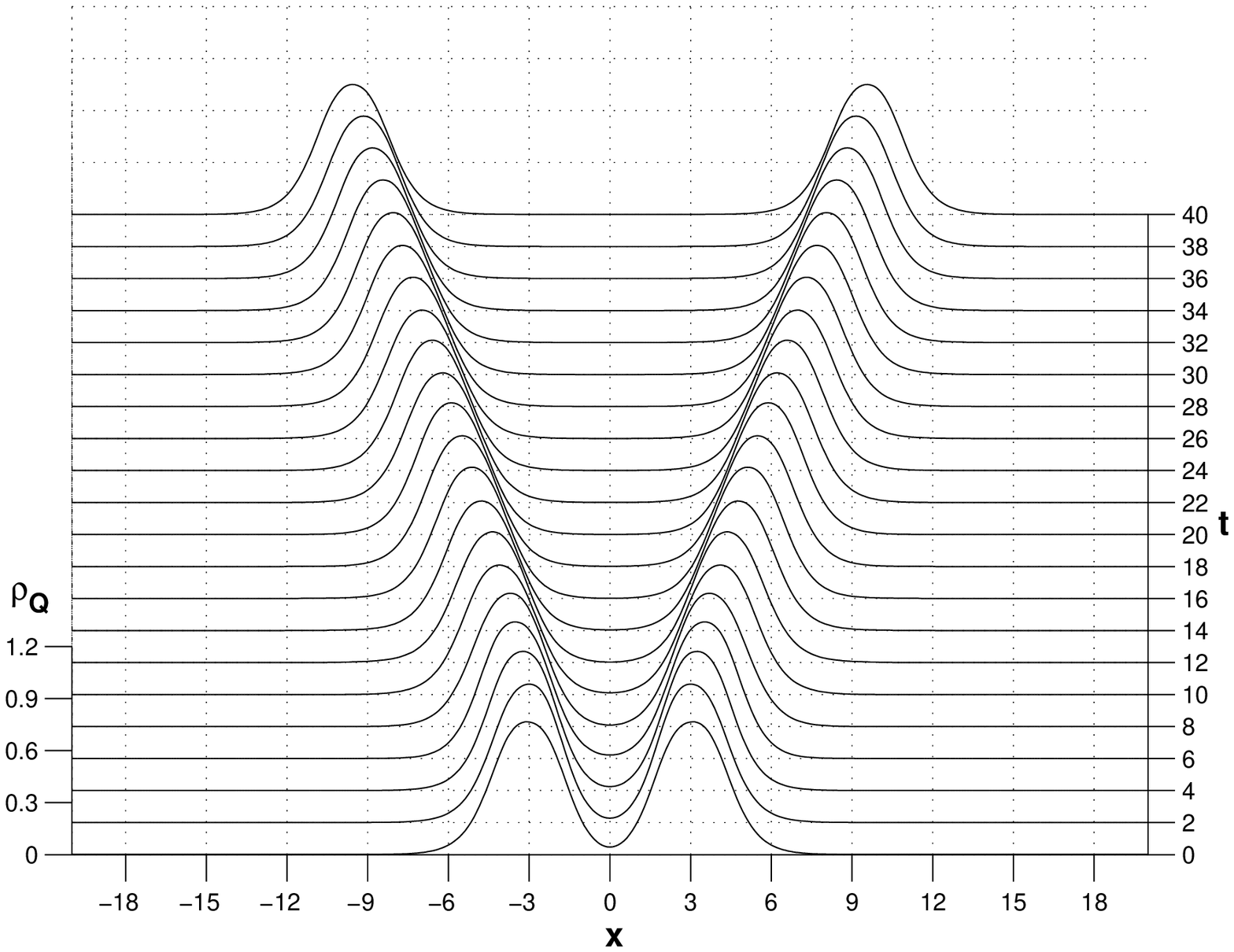}
\includegraphics[width=6.6cm,height=4.5cm]{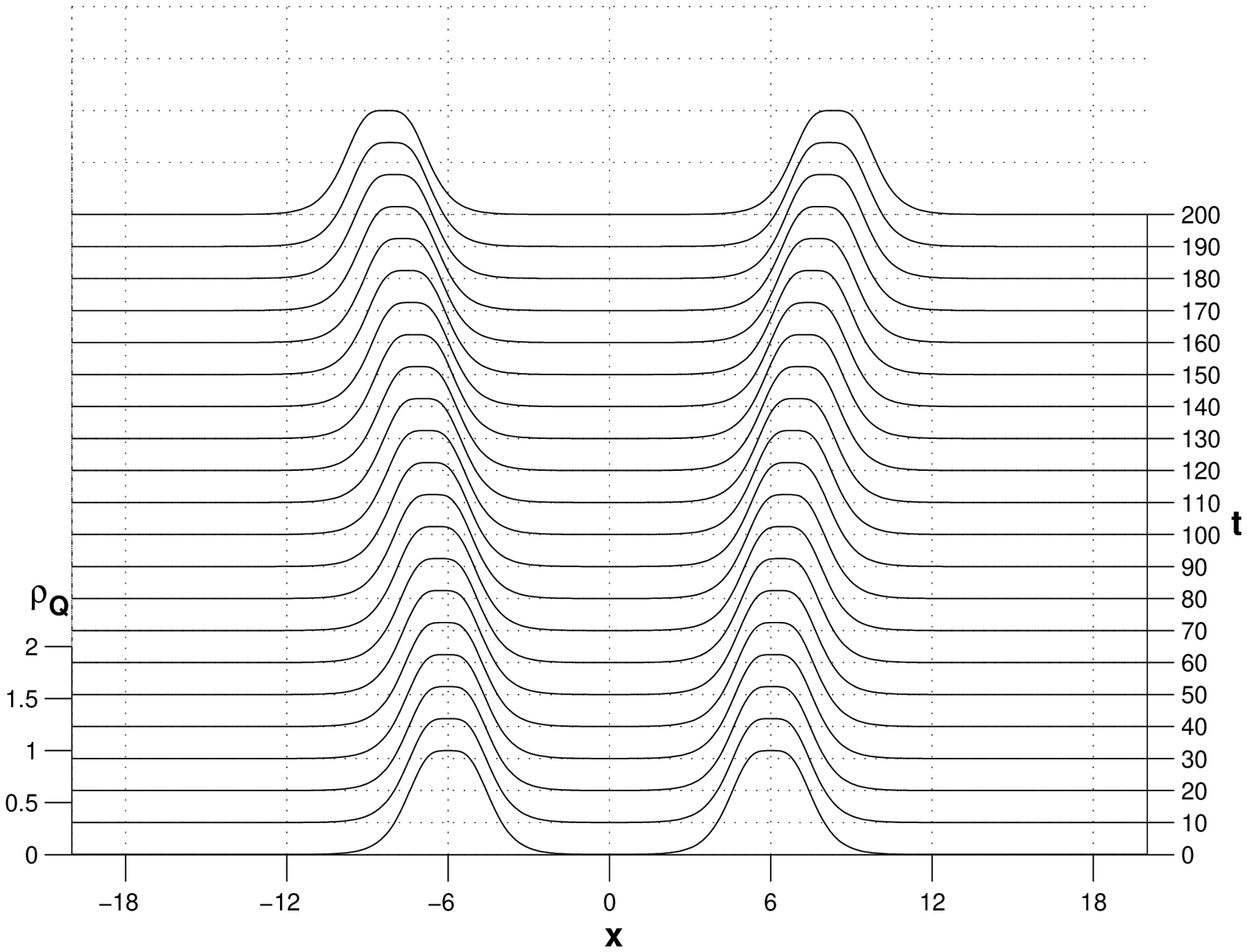}
\caption{\small The time evolution of the charge density $\rho_Q$ for
two initially resting, equal waves.
Left: $\Lambda_l=\Lambda_r=0.6$,  $x_r=-x_l=3$,
 right: $\Lambda_l=\Lambda_r=0.5$,
  $x_r=-x_l=6$.}
\label{fig::caseB2}
\end{figure}

When two  equal waves  are at rest initially and with a phase shift of $\pi$,
 the results given in  Fig.\ \ref{fig::caseB2} show that
 they repulse fully each other. The repulsion force depends on their initial distance.
 The distance is smaller, the  waves move faster in opposition.
 This phenomenon is different from the results on two waves in phase reported in \cite{shao-tang01},
 where the long-lived oscillating bound state is generated when $x_r=-x_l=3$.

\begin{figure}
\includegraphics[width=6.6cm,height=4.5cm]{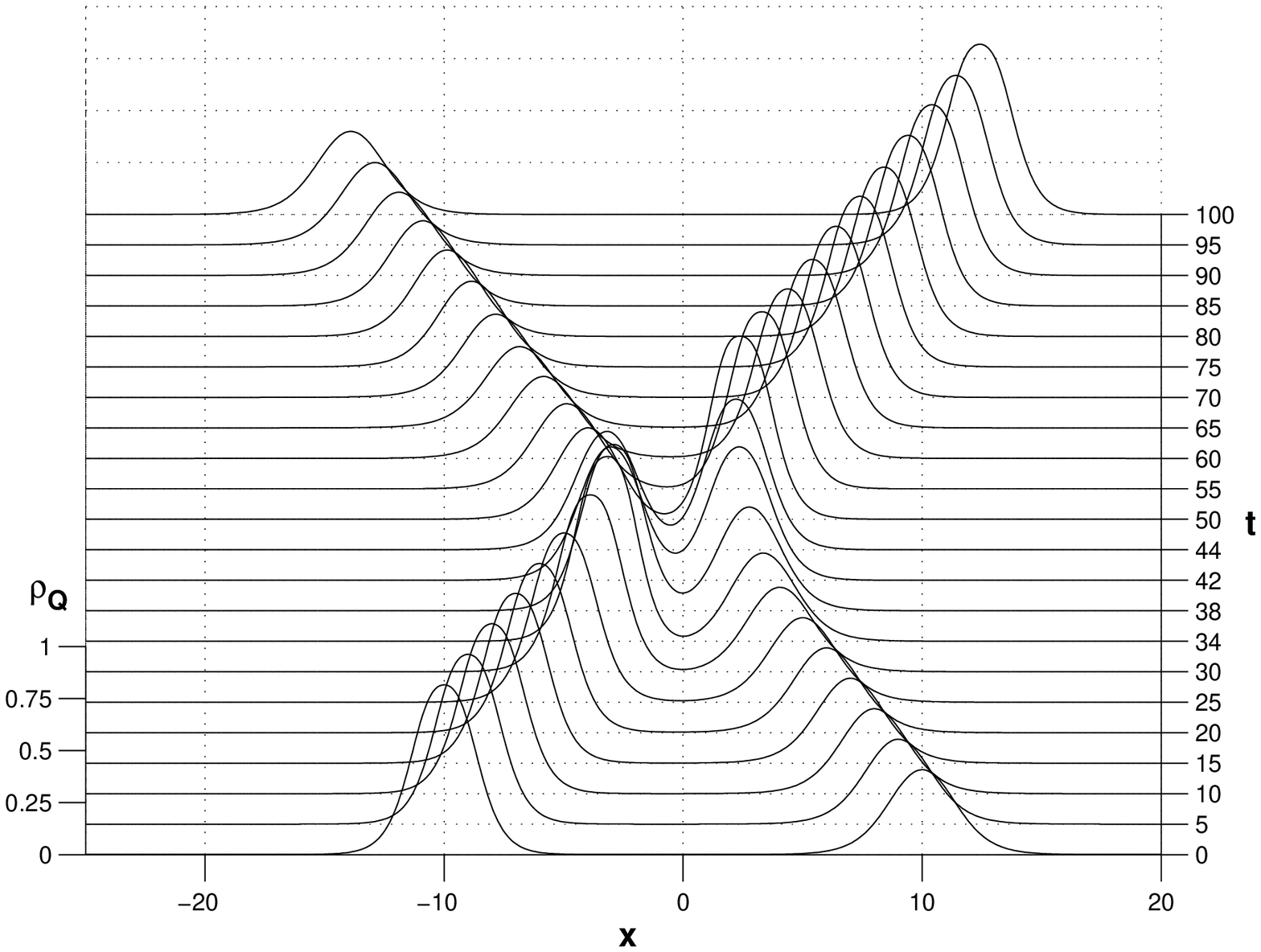}
\includegraphics[width=6.6cm,height=4.5cm]{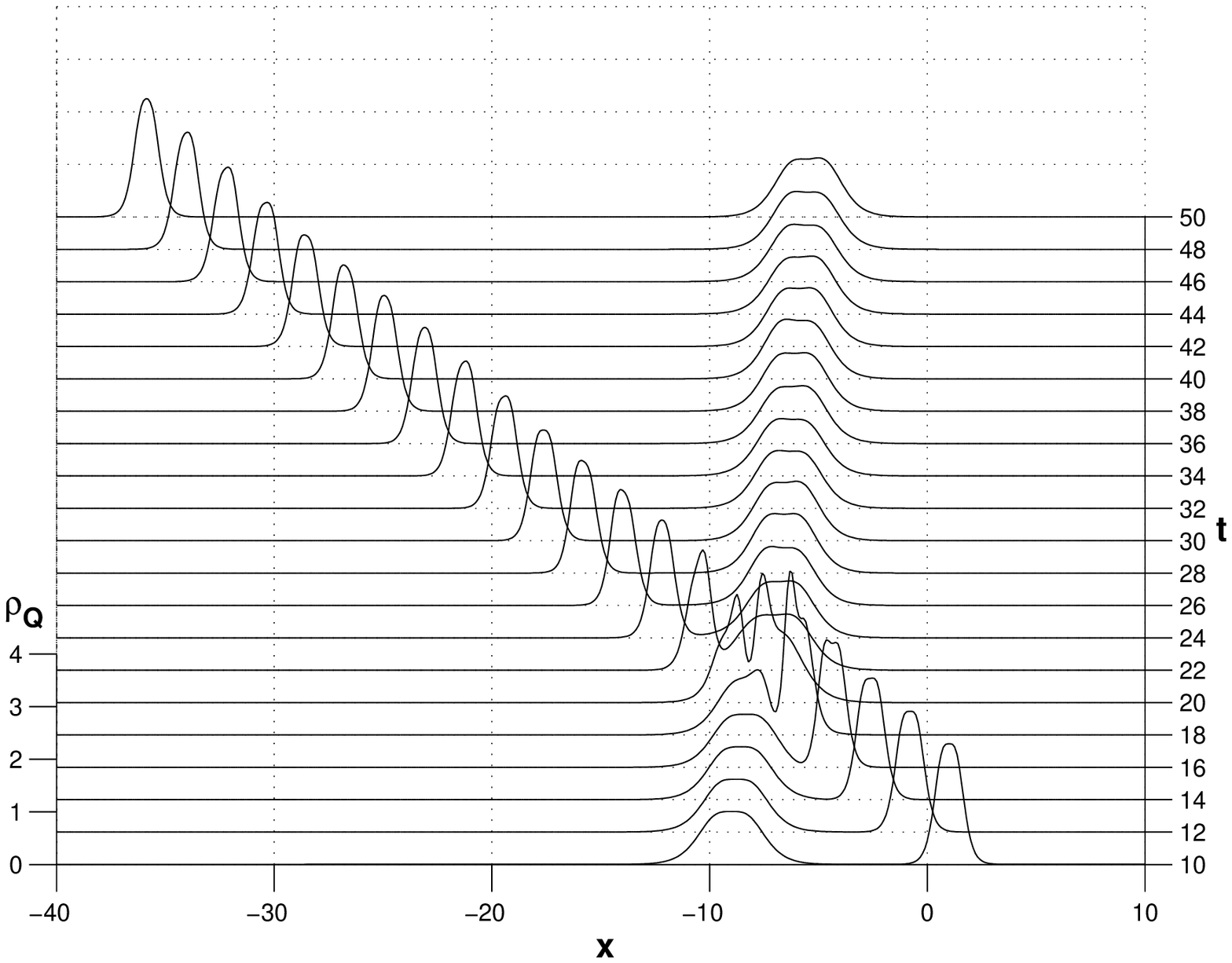}
\caption{\small The time evolution of the charge density $\rho_Q$.
Left: $\Lambda_l=0.6,\Lambda_r=0.8$, $v_l=-v_r=0.2$,
$x_r=-x_l=10$; right:
$\Lambda_l=\Lambda_r=0.5$, $v_l=0.1, v_r=-0.9$, $x_r=-x_l=10$.} \label{fig::caseB3}
\end{figure}

Fig.\ \ref{fig::caseB3} shows the computed results for the case of
two unequal one-humped solitary waves. The left-hand plot is for
the case of $\Lambda_l=0.6, \Lambda_r=0.8$, $v_l=-v_r=0.2$,
and $x_r=-x_l=10$, and the right
figure is for the case of $\Lambda_l=\Lambda_r=0.5$, $v_l=0.1,
v_r=-0.9$,  and $x_r=-x_l=10$.
 We see that the left-hand (or right-hand) initial solitary wave transfers
charge and energy to the right-hand (or left-hand) one and the final
solitary waves are moving with their initial velocities.   Actually,
we may consider that as a traversing or penetration  phenomenon,
that is to say, two waves go through each other in their
interaction.

We have conducted various different experiments on binary
collisions of one-humped waves and
 concluded that in general collapse phenomenon cannot be observed in
 collisions between  two one-humped waves.
To save space, we do not give corresponding plots here.

\subsection{Two  two-humped solitary waves}

This subsection is to study the interaction of two
two-humped solitary waves with a phase shift of $\pi$.

\begin{figure}
\centering
\includegraphics[width=6.6cm,height=4.5cm]{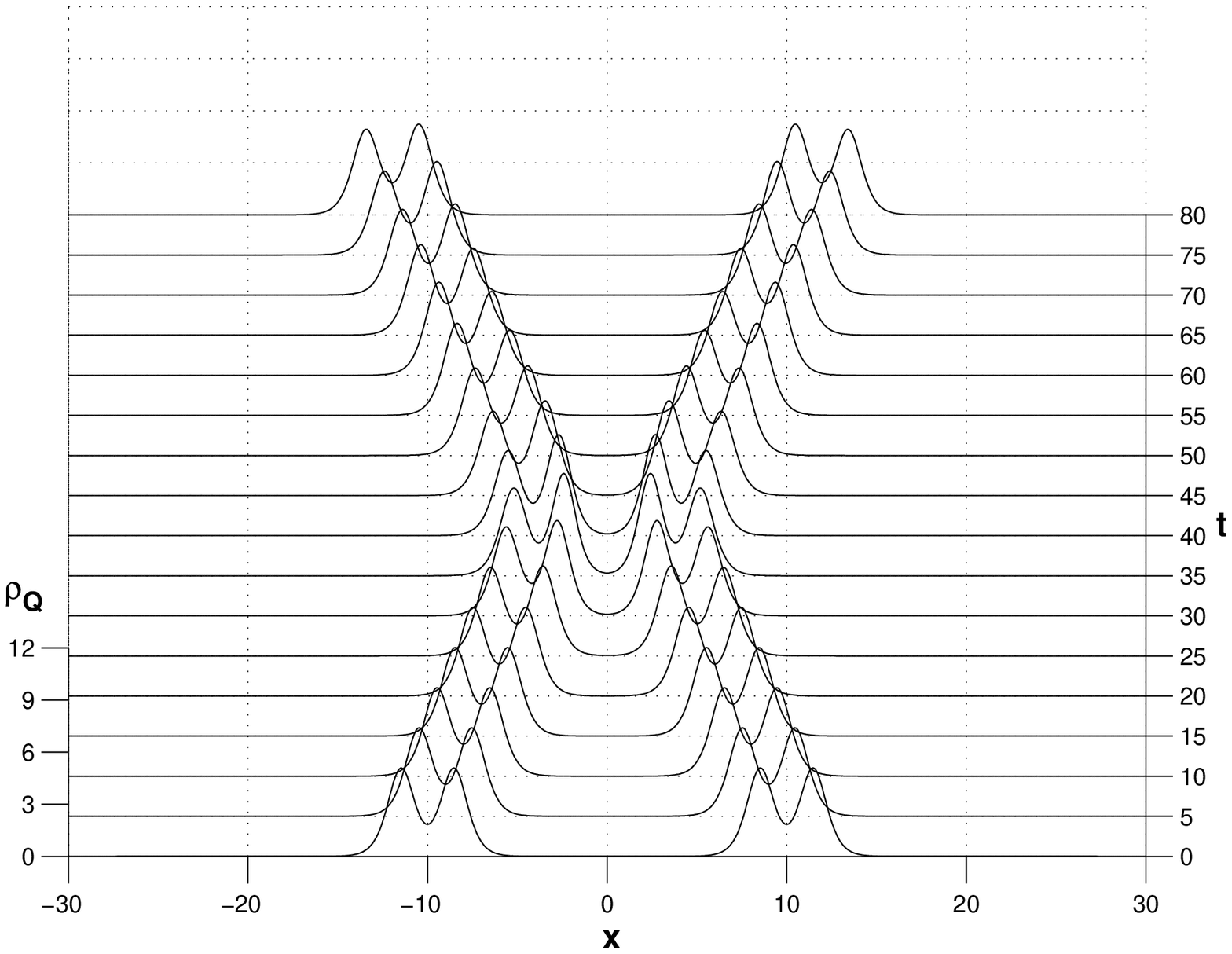}
\includegraphics[width=6.6cm,height=4.5cm]{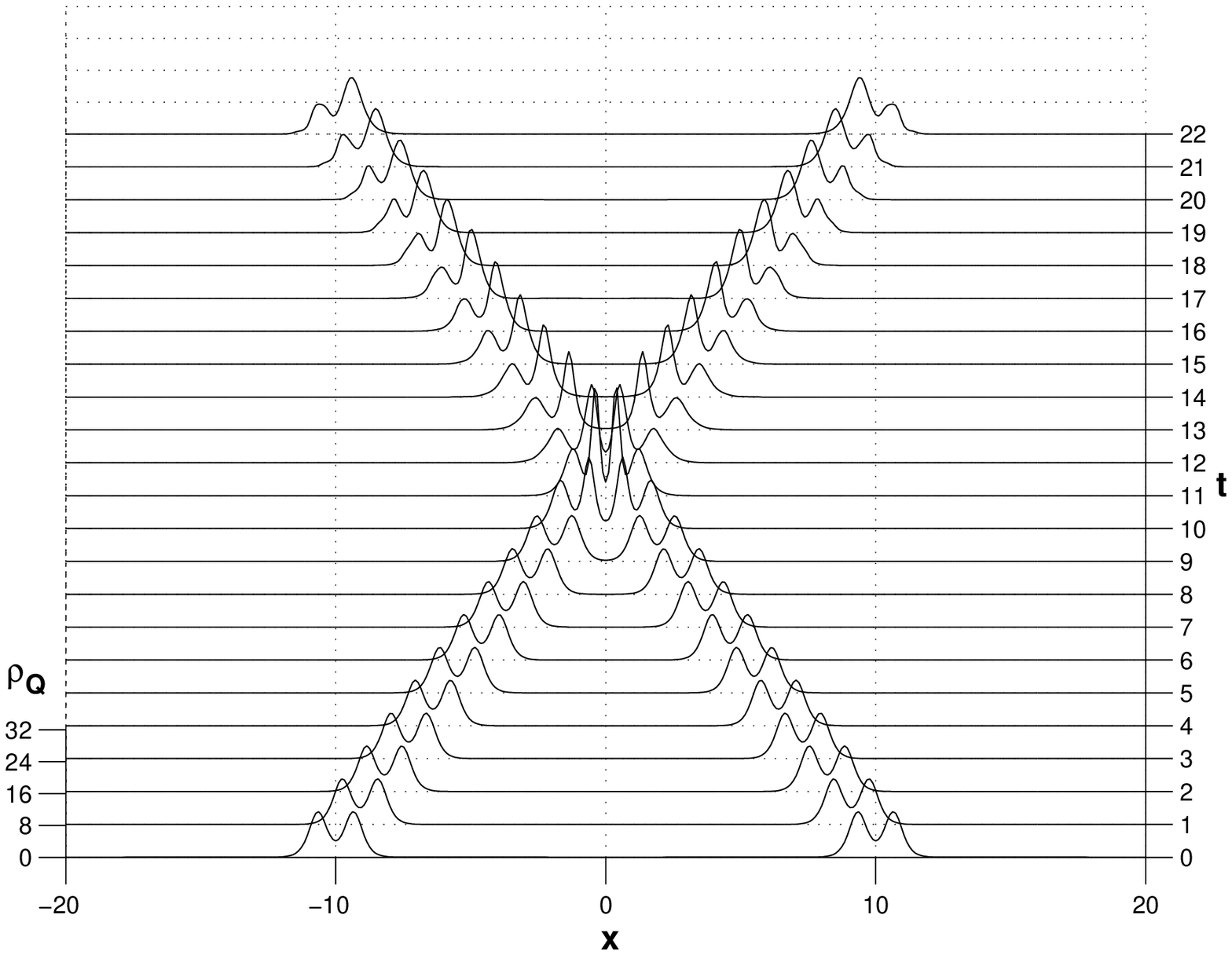}
\caption{\small The time evolution of the charge density $\rho_Q$.
$\Lambda_l=\Lambda_r=0.1$,  $x_r=-x_l=10$. Left: $v_l=-v_r=0.2$; right: $v_l=-v_r=0.9$.}
 \label{fig::caseB4}
\end{figure}

Fig.\ \ref{fig::caseB4} shows the computed results for the cases
of $\Lambda_l=\Lambda_r=0.1$, $v_l=-v_r=0.2$ and $0.9$,
and $x_r=-x_l=10$.
 We observe that the final solitary waves keep their
initial velocities but with  different shapes; the collapse does not
happen in these both cases.
It is worth noting that   the collapse will happen if those two
waves are in phase, see Fig.\ 2 in \cite{shao-tang02}.

\begin{figure}
\includegraphics[width=6.6cm,height=4.5cm]{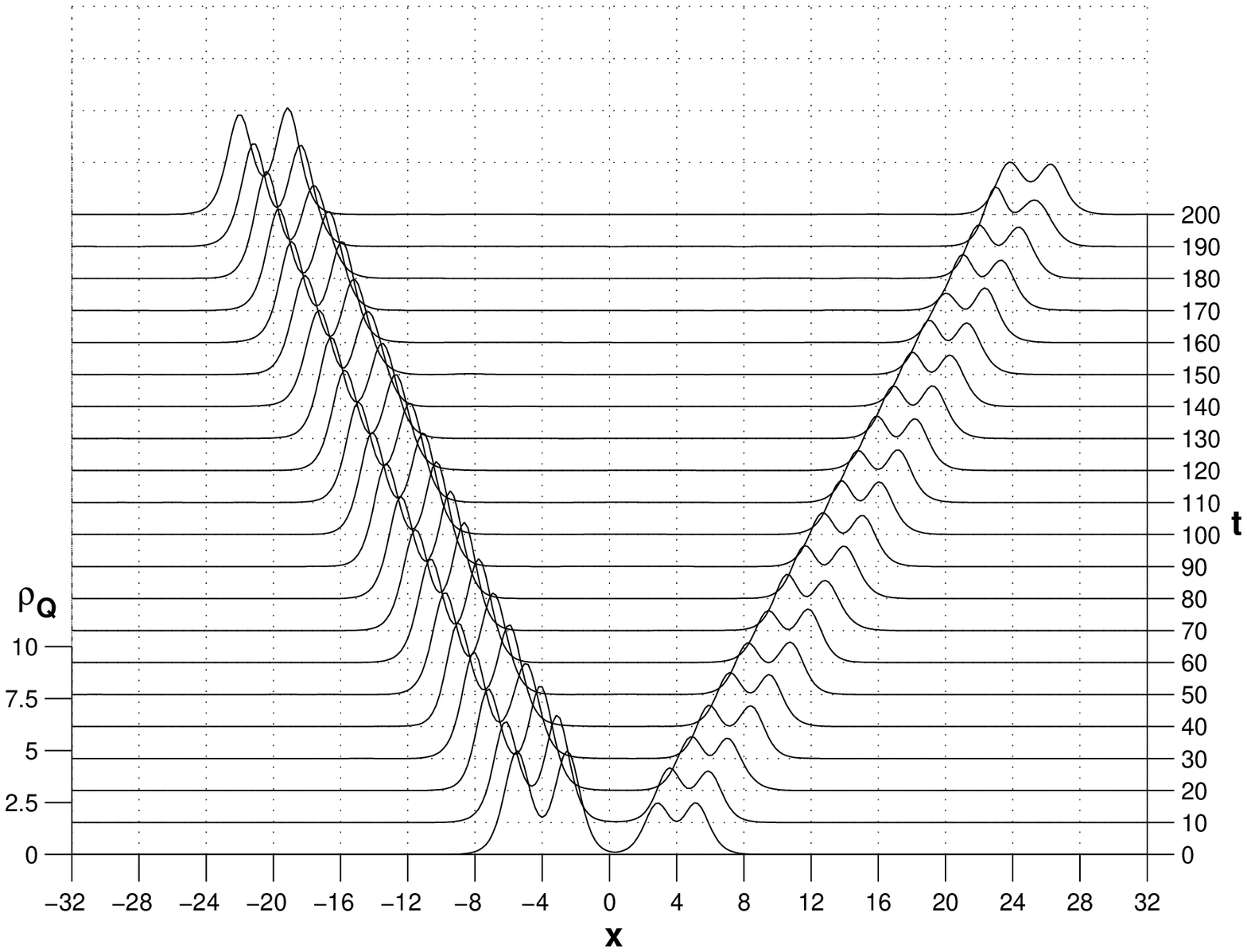}
\includegraphics[width=6.6cm,height=4.5cm]{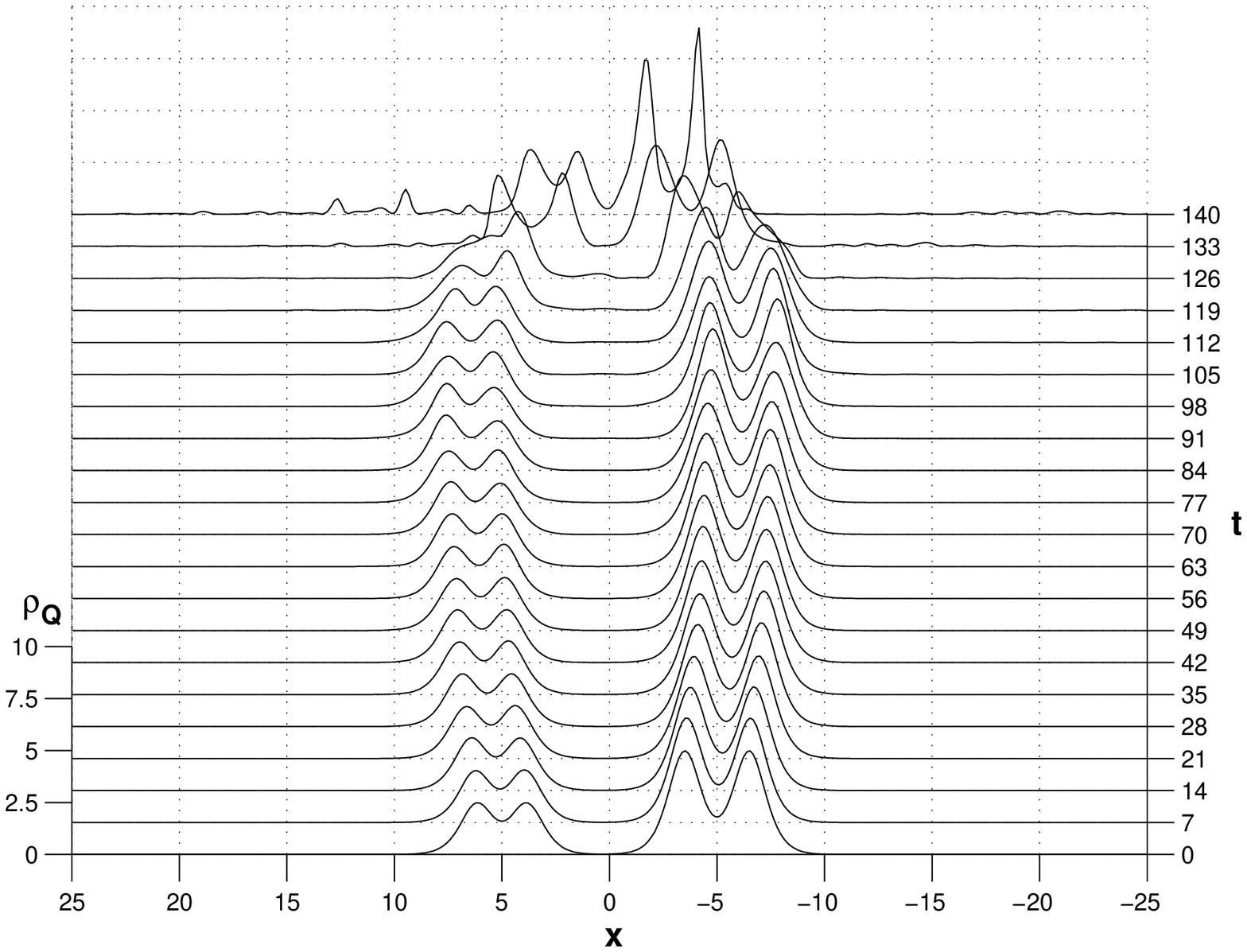}
\caption{\small The time evolution of the charge density $\rho_Q$ for two initially resting waves.
$\Lambda_l=0.1,\Lambda_r=0.2$.
Left: $x_r=-x_l=4$;
right: $x_r=-x_l=5$.}
\label{fig::caseB5}
\end{figure}

Consider the case of that two  unequal waves  are at rest
initially and with a phase shift of $\pi$. The results given in
Fig.\ \ref{fig::caseB5} show that
 they  repulse each other essentially, and the repulsion force depends on their initial distance.
When two initial waves stand more nearly, the repulsion  dominates
in their interaction, thus they move outside fast and cannot
re-collide each other, see the left figure in  Fig.\
\ref{fig::caseB5}. But when the distance is relatively big, the
right plot of  Fig.\ \ref{fig::caseB5} shows that two waves first
repulse each other, then attract afterwards and collapse.
When two initial waves are equal, at rest, and with a phase shift
of $\pi$, we only observed full repulsion  which is  similar to
that in the left figure of Fig.\ \ref{fig::caseB5}.

Besides the above, we also conduct some other experiments and find
that
 collapse happens easily in  collisions
of two in-phase, equal, two-humped waves,
 but it may not appear in collisions  between two in-phase, unequal, two-humped waves, or
two equal  two-humped waves with a phase shift of $\pi$, or  two unequal,
 two-humped waves with a phase shift of $\pi$.

\subsection{A one-humped solitary wave and a two-humped solitary wave}

This subsection is to study the interaction of a  one-humped
solitary wave and a two-humped solitary wave, which are with a
phase shift of $\pi$.

\begin{figure}
\centering
\includegraphics[width=10cm,height=8cm]{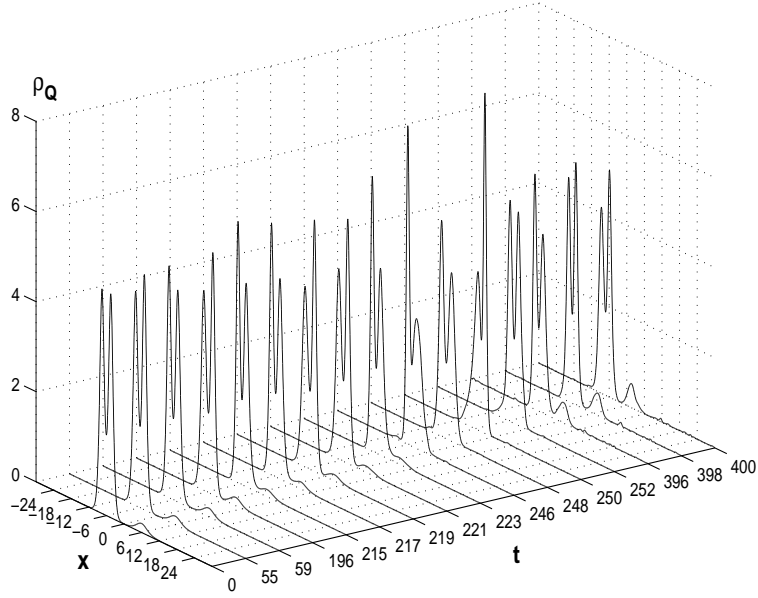}
\caption{\small Bound state formed in binary collisions between two initially resting, out-of-phase waves.
 $\Lambda_l=0.1, \Lambda_r=0.9$, $v_l=v_r=0$, and $x_r=-x_l=6$.} \label{fig::caseB6}
\end{figure}

Fig.\ \ref{fig::caseB6} shows the computed results for the case of
$\Lambda_l=0.1, \Lambda_r=0.9$, $v_l=-v_r=0$, and $x_r=-x_l=6$. We see the quasi-stable
long-lived oscillating bound state, which is essentially same as
one shown in Fig.\ 3 in the paper \cite{shao-tang02}. Generally,
when there is a big difference between the peak values of two
initial waves, macroscopical behavior of the interaction dynamics
are essentially independent on their phase shift.

\begin{figure}
\centering
\includegraphics[width=6.6cm,height=4.5cm]{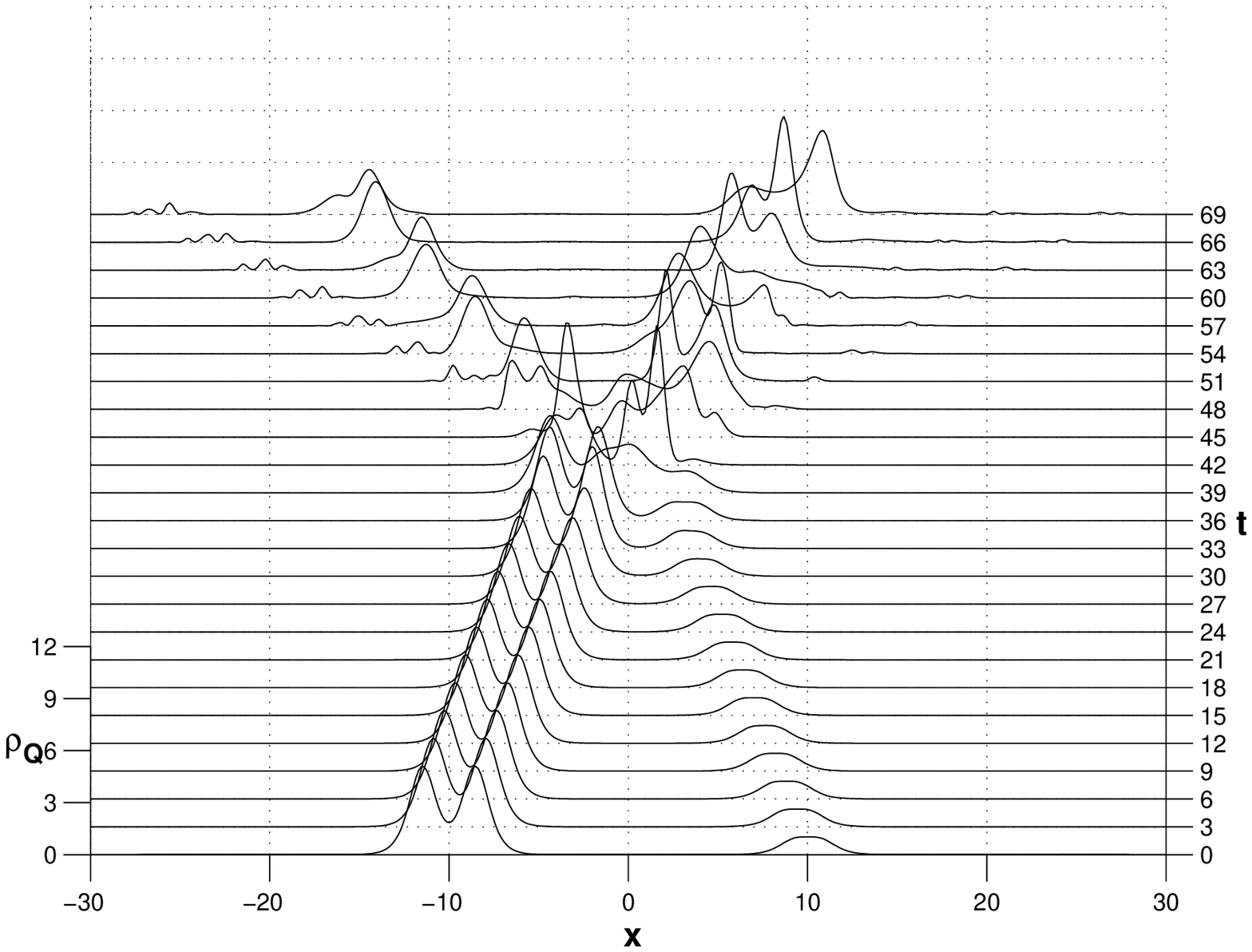}
\includegraphics[width=6.6cm,height=4.5cm]{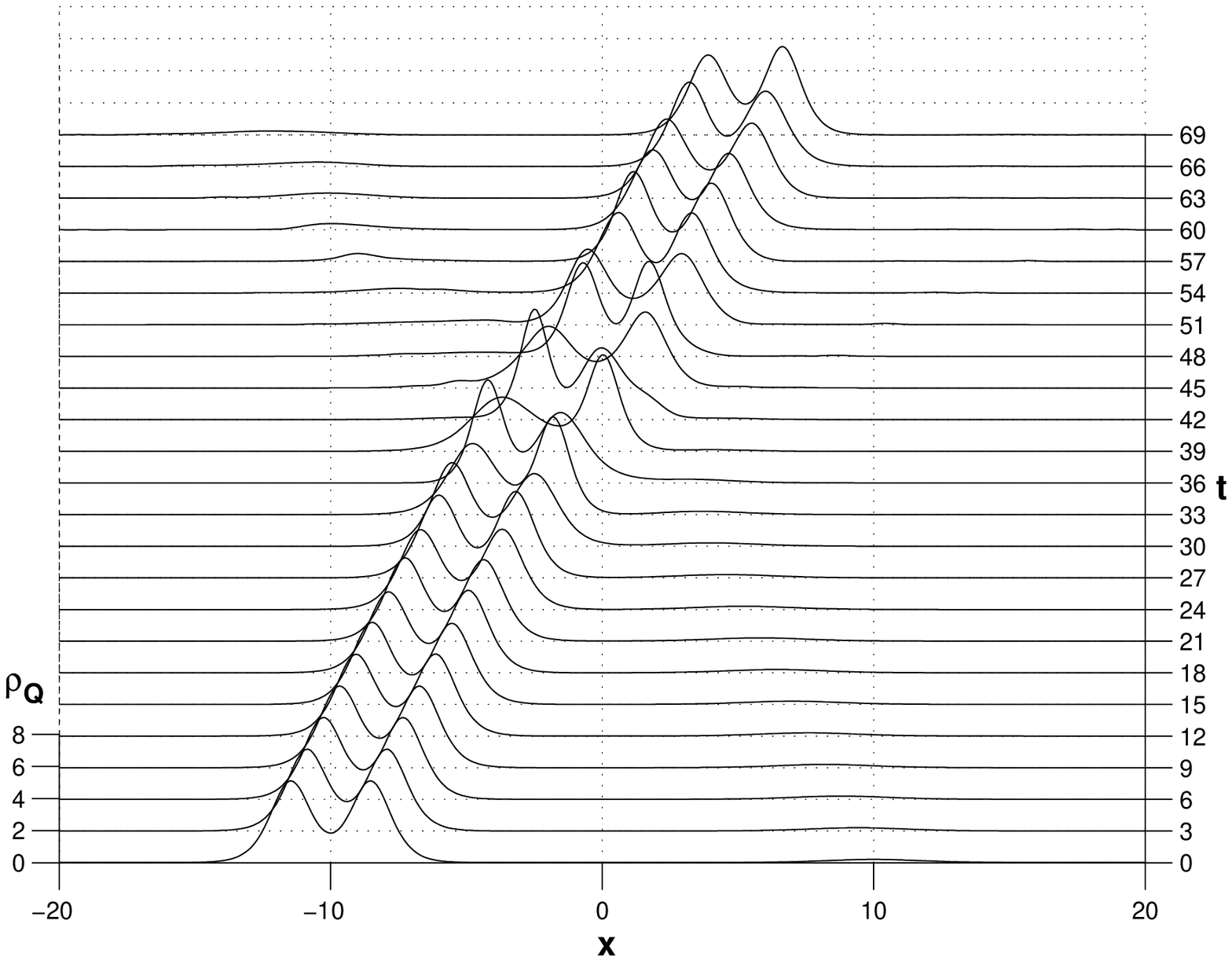}
\caption{\small The time evolution of the charge density $\rho_Q$.
 $\Lambda_l=0.1$, $v_l=-v_r=0.2$, $x_r=-x_l=10$.
 Left:  $\Lambda_r=0.5$; right:  $\Lambda_r=0.9$.} \label{fig::caseB7}
\end{figure}

Fig.\ \ref{fig::caseB7} shows the computed results for the cases
of $\Lambda_l=0.1, \Lambda_r=0.5$ and $0.9$, $v_l=-v_r=0.2$, and
$x_r=-x_l=10$.
It tells us that collapse may be observed in collisions between a
one-humped solitary wave and a two-humped solitary wave
when they do initially travel face to face,
but it may also not  happen.

\section{Ternary collisions}
In this section, we study  ternary collisions by solving the
nonlinear Dirac model (\ref{eq::dirac1}) with the following
initial data
\begin{equation}\label{eq:tecs1}
\vec{\psi}(x,0)=e^{\mi\theta_l}\vec{\psi}^{ss}_l(x-x_l,0)+e^{\mi\theta_m}\vec{\psi}^{ss}_m(x-x_m,0)+e^{\mi\theta_r}
\vec{\psi}^{ss}_r(x-x_r,0),
\end{equation}
where $\theta_l$, $\theta_m$ and $\theta_r$ are three real numbers, determining
the initial phase of corresponding waves.

\begin{figure}
\centering
\includegraphics[width=6.6cm,height=4.5cm]{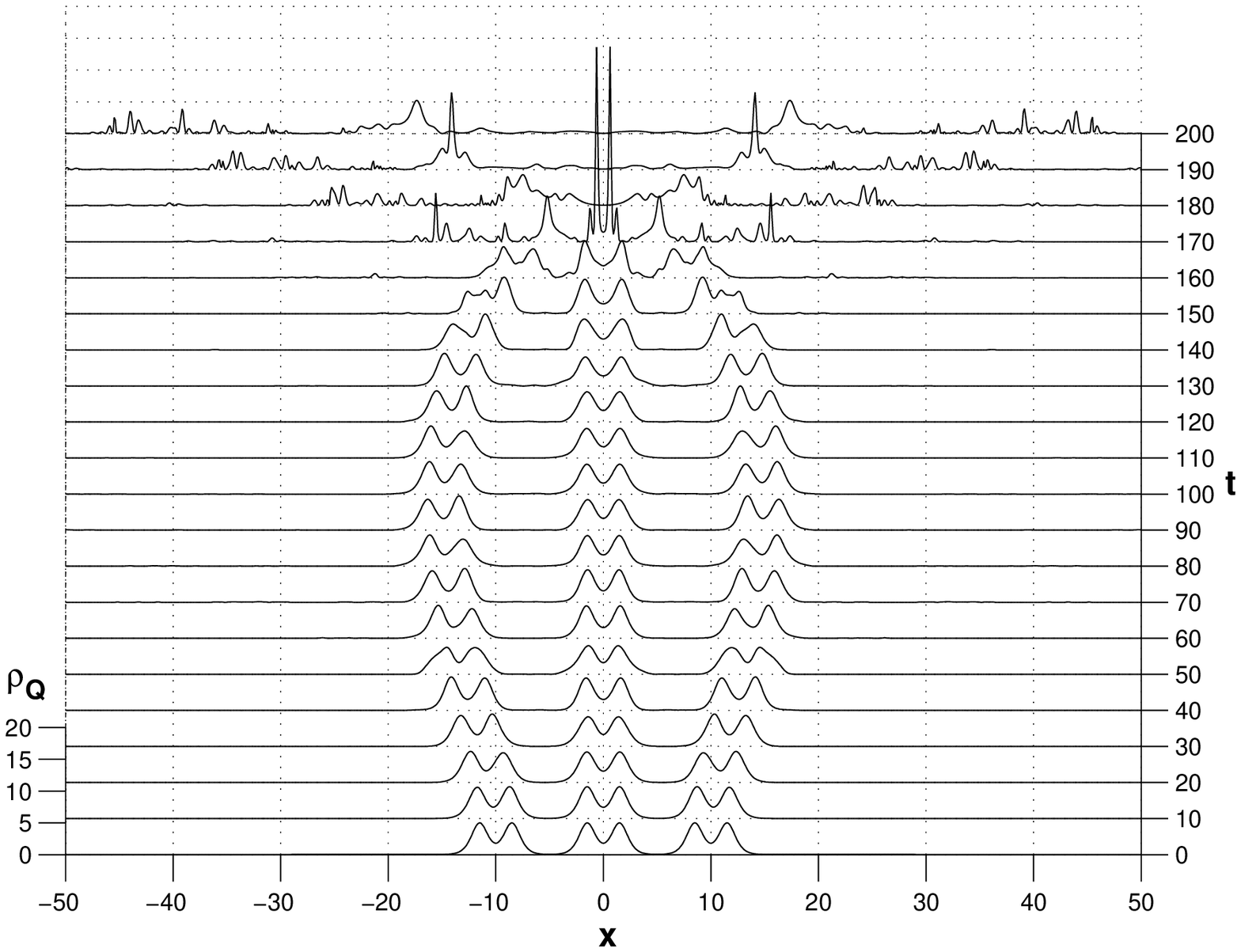}
\includegraphics[width=6.6cm,height=4.5cm]{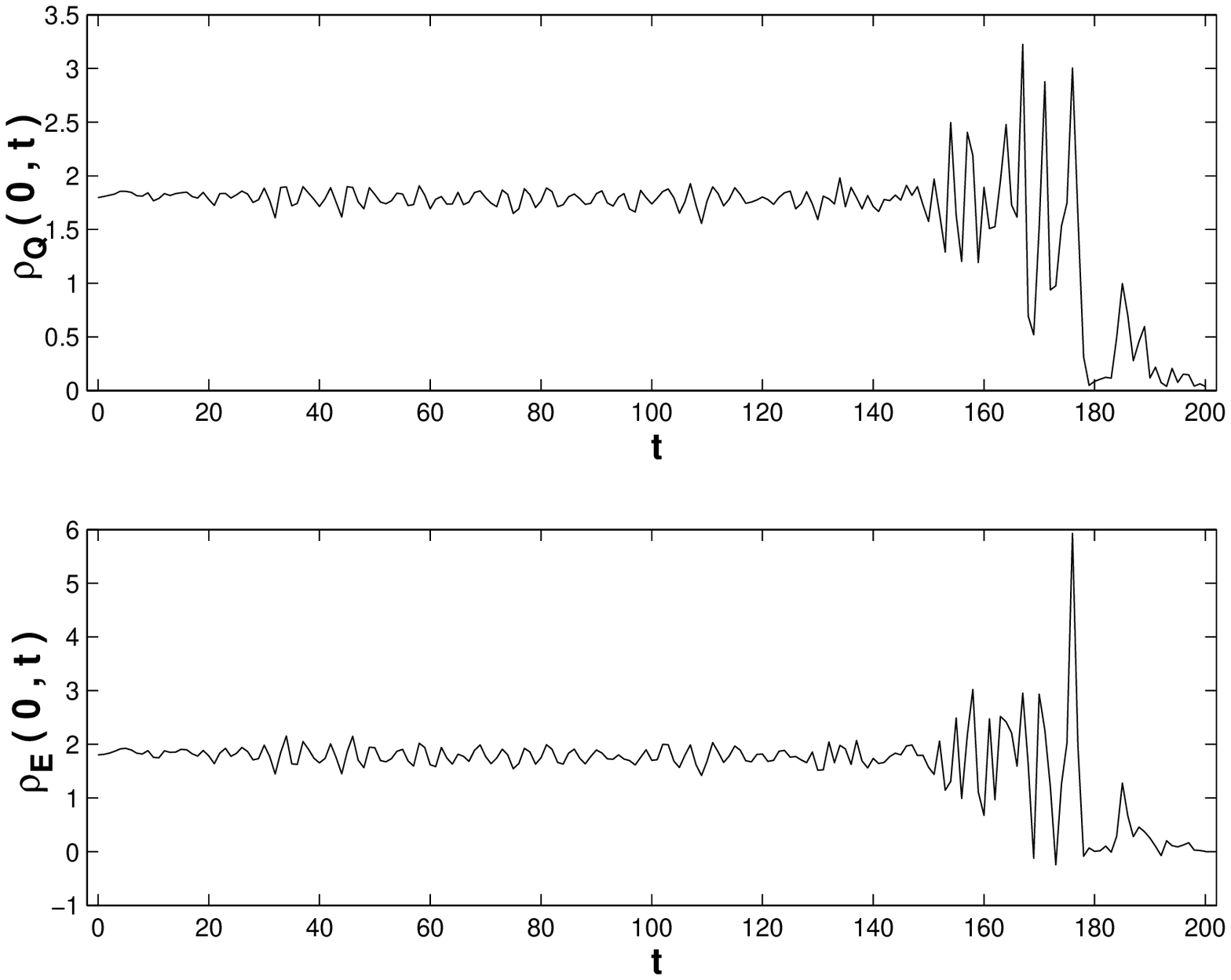}
\caption{\small Collisions of three initially resting two-humped
waves: $\Lambda_l=\Lambda_m=\Lambda_r=0.1$, $x_r=-x_l=10$,
$x_m=0$, $\theta_l=\theta_r=0$, $\theta_m=\pi$. Left:  plot of
$\rho_Q(x,t)$; right: plots of $\rho_Q(0,t)$ and $\rho_E(0,t)$. }
\label{fig::caseT6_1}
\end{figure}

The first case we consider is collisions of three initially
resting two-humped solitary waves:
$\Lambda_l=\Lambda_m=\Lambda_r=0.1$, $v_l=v_m=v_r=0$,
$x_r=-x_l=10$, $x_m=0$, $\theta_l=\theta_r=0$, and $\theta_m=\pi$.
The results are shown in Fig. \ref{fig::caseT6_1}. We see  from
the left plot that the waves first repulse each other because two
neighboring waves are out-of-phase, but after $t=100$ they begin
to attract each other, and final collision results in collapse.
Symmetry of the solutions is kept very well. The right figure of
Fig. \ref{fig::caseT6_1} gives the charge and energy densities at
$x=0$ as a function of time. We observe that before collapse
happens, the middle wave is oscillating because of bind from left
and right waves although its displacement  seems to be unchanged.

\begin{figure}
\centering
\includegraphics[width=6.6cm,height=4.5cm]{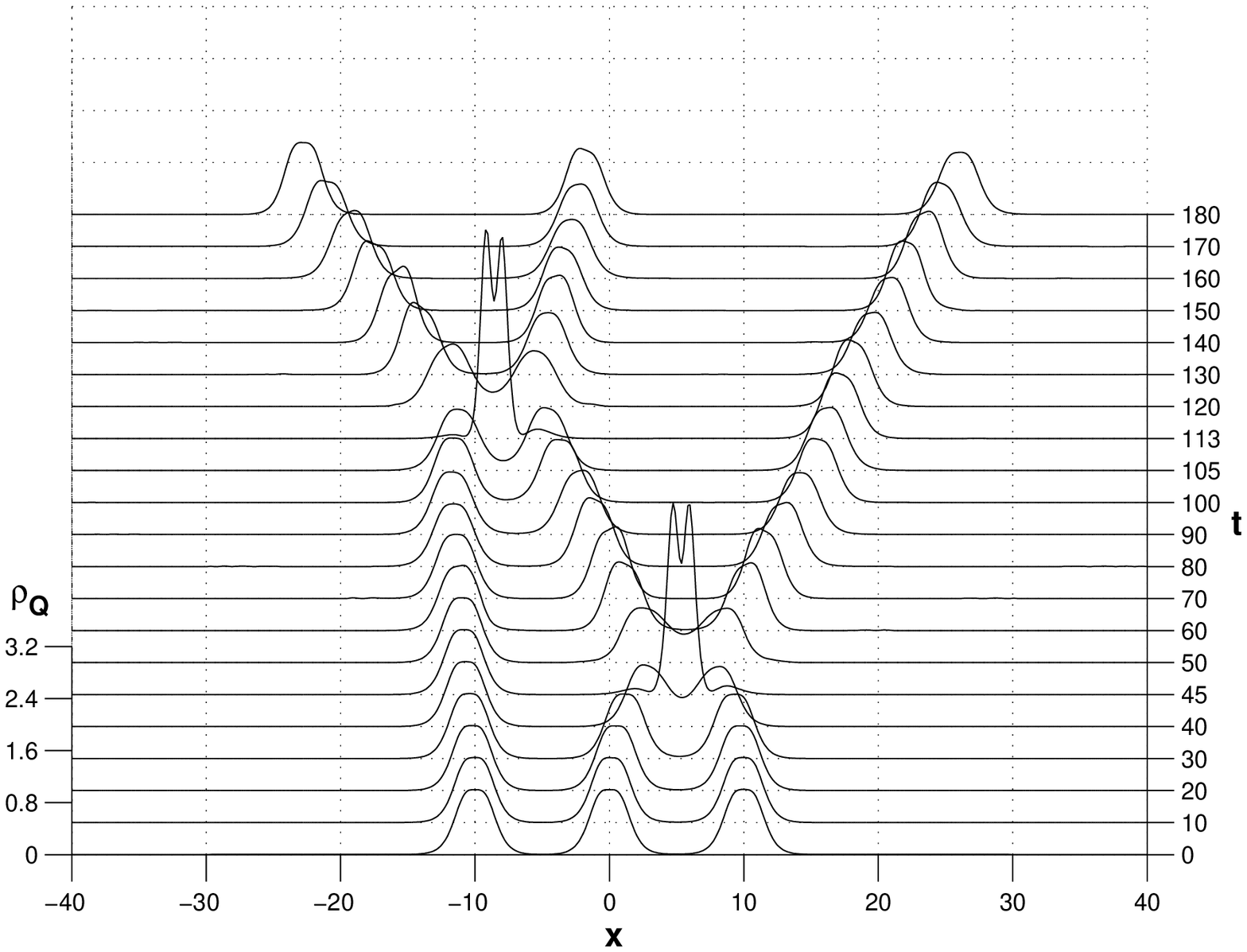}
\includegraphics[width=6.6cm,height=4.5cm]{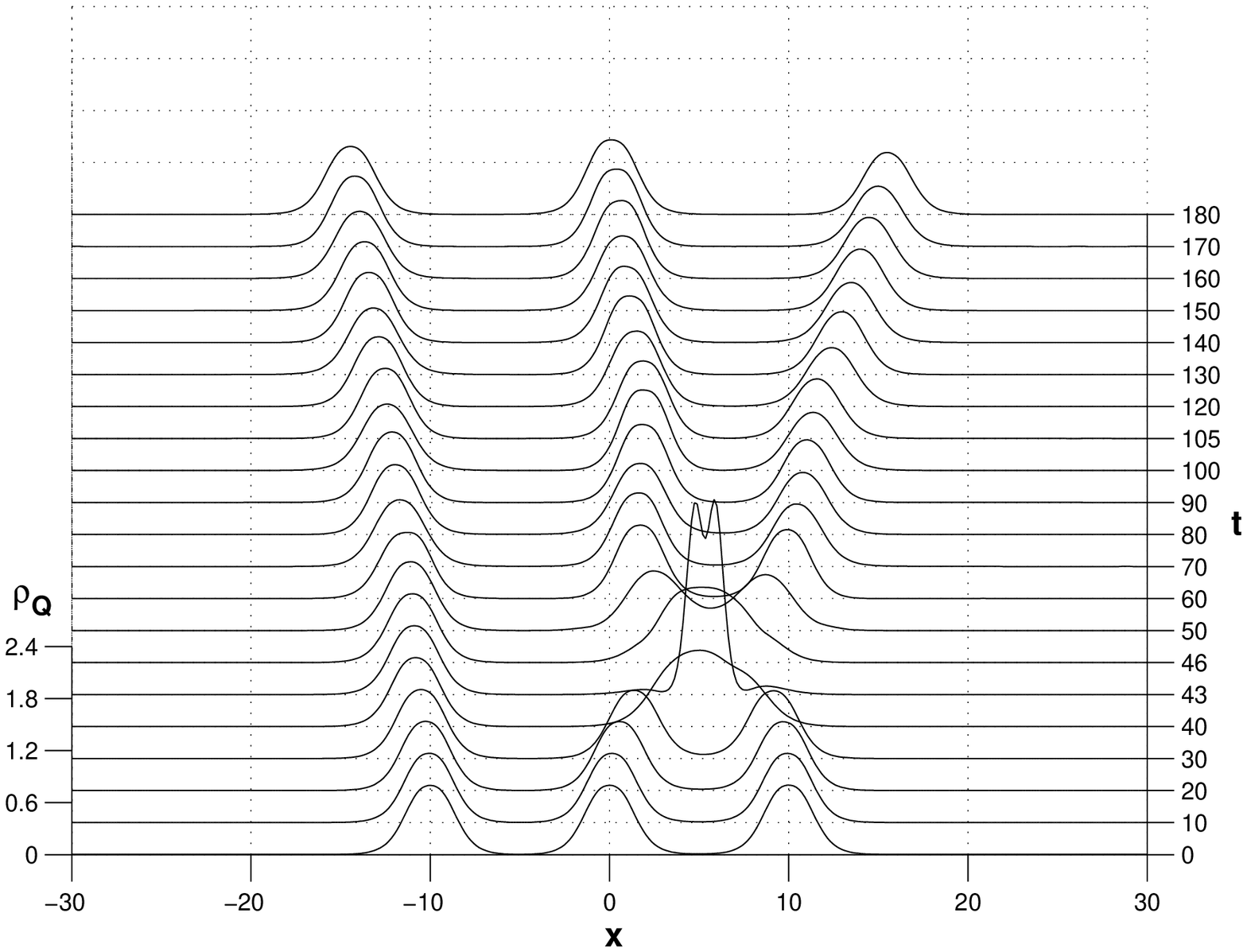}
\caption{\small The time evolution of the charge density $\rho_Q$.
$v_l=v_m=v_r=0$, $x_r=-x_l=10$, $x_m=0$, $\theta_l=\pi$,
$\theta_m=\theta_r=0$. Left: $\Lambda_l=\Lambda_m=\Lambda_r=0.5$;
right: $\Lambda_l=\Lambda_m=\Lambda_r=0.6$.} \label{fig::caseT8}
\end{figure}

The second case is collisions of three initially resting
one-humped waves: $\Lambda_l=\Lambda_m=\Lambda_r=0.5$,
$v_l=v_m=v_r=0$, $x_r=-x_l=10$, $x_m=0$, $\theta_l=\pi$, and
$\theta_m=\theta_r=0$. The results are shown in the left plot of
Fig. \ref{fig::caseT8}. We see that
 the first interaction happens between two right in-phase waves around $t=45$, and
then  two faster moving waves are formed. The initially resting left
wave begin to be moving towards left due to repulsion between it and
the right waves, and then it is catched up with and interacted by
the left moving wave generated in the first interaction around
$t=130$. Overlapping happens in all two interactions. If we consider
collisions of three initially resting one-humped waves with
$\Lambda_l=\Lambda_m=\Lambda_r=0.6$ or 0.9,
the interaction does only happen between two right neighboring
in-phase waves, see the right plot of Fig. \ref{fig::caseT8}. The
reason is that the left-moving (middle) wave formed in the
interaction of two initially in-phase (right) waves is not faster
than the left wave.

\section{Discussions and Conclusions}
In this paper, we further studied the interaction dynamics for the
solitary waves of a nonlinear Dirac model (\ref{eq::dirac1}) with
scalar self-interaction. By using a fourth--order accurate RKDG
method presented in \cite{shao-tang01}, we investigated the
interaction of the Dirac solitary waves with an initial phase
shift of $\pi$ for the first time, and observed that:
 (a) full repulsion in binary and ternary collisions, and the initial distance between waves
 is smaller the repulsion is stronger; (b) repulsing first,
attracting afterwards, and then collapse
 in binary and ternary collisions of two-humped waves; (c) interaction with one
 overlap and two overlaps in ternary collisions of initially resting waves, which depends on
 initial parameter $\Lambda$. We concluded that in general collapse phenomenon cannot be observed in
 collisions between one-humped waves, but it happens easily in  collisions
of in-phase, equal, two-humped waves;  the macroscopical behavior
of the interaction dynamics is essentially independent on their
initial phase difference, when there is a big difference between
the peak values of initial waves. Although we have investigated
the influence of initial phase difference on the interaction
dynamics of the Dirac solitary waves, it will be   interesting and
important  to study the evolution of the relative phase of the
solitons during their collisions.

\section*{Acknowledgments}
This research was   partially supported by
 the National Basic Research Program under the Grant 2005CB321703,
 the National Natural Science Foundation of China (No.\ 10431050, 10576001),
  and Laboratory of Computational Physics.


\begin{thebibliography}{99}
\small

\bibitem{soler1970} M. Soler,
Classical, stable, nonlinear spinor field with positive rest energy,
 Phys. Rev. D 1(1970), 2766-2769.


\bibitem{rs1973} A.F. Ra$\tilde{\text{n}}$ada, M. Soler,
Perturbation theory for an exactly soluble spinor model in
interaction with its electromagnetic field,
Phys. Rev. D 8(1973), 3430-3433.


\bibitem{rrsv1974} A.F. Ra$\tilde{\text{n}}$ada, M.F.
Ra$\tilde{\text{n}}$ada, M. Soler, L. V$\acute{\text{a}}$zquez,
Classical electrodynamics of a nonlinear Dirac field with anomalous magnetic moment,
Phys. Rev. D 10(1974), 517-525.

\bibitem{alvarez1983ene} A. Alvarez, M. Soler,
Energetic stability criterion for a nonlinear spinorial model,
Phys. Rev. Lett. 50(1983), 1230-1233.

\bibitem{as1986} A. Alvarez, M. Soler, Stability of the minimum solitary wave of a nonlinear spinorial model,
Phys. Rev. D 34(1986), 644-645.


\bibitem{alvarez1985} A. Alvarez, Spinorial solitary wave dynamics of a (1+3)-dimensional model,
Phys. Rev. D 31(1985), 2701-2703.

\bibitem{alvarez1988}
A. Alvarez and A. F. Ra$\tilde{\mbox{n}}$da,
Blow-up in nonlinear models of extended particles with confined constituents,
Phys. Rev. D 38(1988), 3330-3333.


\bibitem{alvarez1981} A. Alvarez, B. Carreras, Interaction dynamics for the solitary waves of a nonlinear Dirac model,
Phys. Lett. A 86(1981), 327-332.



\bibitem{alvarez1992} A. Alvarez, Linearized Crank-Nicholson scheme for nonlinear Dirac equations,
 J. Comput. Phys. 99(1992), 348-350.

\bibitem{sss} J. De Frutos, J.M. Sanz-serna, Split-step spectral schemes for nonlinear Dirac systems,
J. Comput. Phys. 83(1989), 407-423.


\bibitem{wg2004} Z.-Q. Wang, B.-Y. Guo, Modified Legendre rational spectral method
for the whole line, J. Comput. Math. 22(2004),
457-474.



\bibitem{shao-tang01} S.H. Shao, H.Z. Tang, Higher-order accurate Runge-Kutta
discontinuous Galerkin methods for a nonlinear Dirac model,
Discrete and continuous dynamical systems-series B 6(2006),
623-640.


\bibitem{shao-tang02}
S.H. Shao, H.Z. Tang, Interaction for the solitary waves of a
nonlinear Dirac model,  Phys. Lett. A, 345(2005), pp.119-128.

\end{thebibliography}
\end{document}